# Cosmology, initial conditions, and the measurement problem

David Layzer
Department of Astronomy, Harvard University, 60 Garden Street, Cambridge, MA 02478

The assumption that a *complete* description of an early state of the universe does not privilege any position or direction in space leads to a unified account of probability in cosmology, macroscopic physics, and quantum mechanics. Such a description has a statistical character. Deterministic laws link it to statistical descriptions of the cosmic medium at later times, and because these laws do not privilege any position or direction in space, the same must be true of these descriptions. If the universe is infinite, we can identify the probability that the energy density at a particular instant and a particular point in space (relative to a system of spacetime coordinates in which the postulated spatial symmetries are manifest) lies in a given range with the fractional volume occupied by points where the energy density lies in this range; and similarly with all other probabilities that figure in the statistical description. The probabilities that figure in a complete description of the cosmic medium at any given moment thus have an exact and objective physical interpretation. The statistical entropy and the information associated with each cosmological probability distribution are likewise objective properties of the universe, defined in terms of relative frequencies or spatial averages.

The initial states of macroscopic systems are characterized by probability distributions of microstates, linked by a deterministic historical account to probability distributions that characterize the early universe. The probability distributions that characterize the initial states of macroscopic systems are usually deficient in particular kinds of microscopic information -- though the history of a macroscopic system may include an experimental intervention that creates a hidden form of microscopic information, as in Hahn's spin-echo experiment. These general conclusions clarify the relation between quantum mechanics and quantum statistical mechanics. A macroscopic system's history determines the statistical ensembles that represent its macrostates and, absent experimental interventions that produce hidden microscopic order, justifies what van Kampen has called the "repeated randomness assumptions" needed to derive master equations and *H* theorems.

I give a schematic account of quantum measurement that combines von Neumann's definition of an ideal measurement and his account of "premeasurements" with the conclusion that the initial state of a measuring apparatus is fully characterized by a historically determined probability distribution of microstates that is deficient in microscopic information. To comply with the postulated requirement that physical descriptions must not privilege our (or any other) position, I interpret the post-measurement probability distribution of quantum states of the combined system as describing the state of an ensemble of identical measuring experiments uniformly scattered throughout the universe. The resulting account predicts (there is no need to invoke von Neumann's collapse postulate) that each apparatus in the cosmological ensemble registers a definite outcome and leaves the quantum system in the corresponding eigenstate of the measured quantity. Because the present account conflates the indeterminacy of measurement outcomes with the indeterminacy of the combined system's position, it reconciles the indeterminacy of quantum measurement outcomes with the deterministic character of Einstein's field equations, thereby invalidating a standard argument for the need to quantize general relativity. A more detailed account of measurement would allow for the effects of decoherence, but as Joos and Zeh, Schlosshauer, and others have emphasized, decoherence theory alone does not explain why quantum measurements have definite outcomes.

Finally, because the initial states of actual macroscopic systems may be objectively deficient in certain kinds of microscopic information, macroscopic processes that are sensitive to initial conditions may have *objectively* indeterminate outcomes. In such processes, as in quantum measurements, a probability distribution that defines a single initial macrostate may evolve (deterministically) into a probability distribution that assigns finite probabilities to two or more macrostates. Examples include chaotic orbits in celestial mechanics, evolving weather systems, and processes that generate diversity in such biological processes as biological evolution and the immune response.

## I. INTRODUCTION

Initial conditions characterize systems to which given physical laws apply and the conditions under which they apply. Wigner [1] has suggested that a "minimal set of initial conditions not only does not permit any exact relations between its elements; on the contrary, there is reason to contend that these are, or at some time have been, as random as the externally imposed gross constraints allow." As an illustration of such initial conditions, he cites a version of Laplace's nebular hypothesis, in which a structureless, spinning gas cloud evolves into a collection of planets revolving in the same sense around a central star in nearly circular, nearly coplanar orbits. The gross constraints in this example include the cloud's initial mass, angular momentum, and chemical composition. "More generally [Wigner writes], one tries to deduce almost all 'organized motion,' even the existence of life, in a similar fashion." This paper seeks to ground this view of initial conditions in a historical account that links the initial conditions that characterize physical systems and their environments to the initial conditions that characterize a statistically uniform and isotropic model of the universe at a time shortly after the beginning of the cosmic expansion.

In such a universe, particle reaction rates exceed the cosmic expansion rate at sufficiently early times [2]. Consequently, local thermal equilibrium prevails at the instantaneous and rapidly changing values of temperature and mass density. Suppose that there is a system of spacetime coordinates relative to which no statistical property of the cosmic medium defines a preferred position or direction. Then the values of a small number of physical parameters, notably the photon-baryon and lepton-baryon ratios [3], suffice to determine the state of the cosmic medium at early times. This paper explores some consequences of the assumption that this characterization of the early cosmic medium is exact and complete – that it contains *only* statistical information and is invariant under spatial translations and rotations. For example, a uniform, unbounded distribution of elementary particles in thermal equilibrium is completely characterized by its temperature and mass density.

The following discussion does not seek to extend quantum mechanics or general relativity beyond their current domains of validity. Rather, it seeks to show how these domains merge smoothly with the macroscopic domain of statistical physics. I argue that macroscopic systems that interact sufficiently weakly with their surroundings can be assigned definite macrostates characterized by probability distributions of microstates that have evolved from probability distributions that characterized the cosmic medium early on. By virtue of a strong version of the cosmological principle, discussed in the next section, these probability distributions represent objective indeterminacy rather than ignorance. The same strong version of the cosmological principle requires one to interpret the classical fields that figure in relativistic cosmology as random fields, just as in the theory of homogeneous turbulence (except that in the cosmological context the probability distributions that characterize the random fields represent objective indeterminacy). I will argue (in section III.E) that this way of indirectly linking quantum mechanics to general relativity enables one to circumvent a standard argument for the need to quantize gravity. Because the argument in section III.E assumes conditions in which quantum mechanics and relativistic cosmology are both valid, it doesn't bear directly on the need for a theory that applies under more extreme conditions (the very early universe, black holes). On the other hand, an account of initial conditions that serves to reconcile quantum mechanics and general relativity under present conditions could conceivably prove to be relevant to the much larger project of unifying the two theories.

## II. THE STRONG COSMOLOGICAL PRINCIPLE

The cosmological principle states that there exists a system of spacetime coordinates relative to which no statistical property of the physical universe defines a preferred position or direction in space. It embraces Hubble's "principle of uniformity," which applies to the observable distribution of galaxies, and Einstein's model of the universe as a uniform, unbounded distribution of dust [35]. Like the assumptions that define idealized models of stars and galaxies, it has usually been viewed as a convenient simplification. Unlike those assumptions, however, the cosmological principle *could* hold exactly. I will refer to the hypothesis that it does hold exactly as the strong cosmological principle [4], [5]. It draws support from the following considerations.

First, precise and extensive observations of the cosmic microwave background and of the spatial distribution and line-of-sight velocities of galaxies have so far produced no evidence of deviations from statistical homogeneity and isotropy. Astronomical observations provide no positive support for the view that the cosmological principle is merely an approximation or an idealization, like the initial conditions that define models of stars and galaxies.

Second, the initial conditions that define the universe do not have the same function as those that define models of astronomical systems. A theory of stellar structure must apply to a range of stellar models because stars have a wide range of masses, chemical compositions, spins, and ages. No analogous requirement obtains for models of the universe.

Finally, general relativity predicts that local reference frames that are unaccelerated relative to the cosmological coordinate system defined by the cosmological principle are inertial [4]. Astronomical evidence supports this prediction. It indicates that local inertial reference frames are indeed unaccelerated relative to a coordinate system in which the cosmic microwave background is equally bright, on average, in all directions and in which the spatial distribution of galaxies is statistically homogeneous and isotropic. If the distribution of energy and momentum on cosmological scales were not statistically homogeneous and isotropic, there would be no preferred cosmological frame and hence no obvious explanation for the observed relation between local inertial frames and the frame defined by the cosmic microwave background and the spatial distribution and line-of-sight velocities of galaxies.

Cosmological models that conform to the cosmological principle are characterized by parameters, such as the photon-to-baryon and lepton-baryon ratios, the curvature scale, and the cosmological constant. Some theories for the origin of structure also posit primordial density fluctuations whose spectrum and amplitude are characterized by parameters that do not privilege particular spatial positions or directions. Because current physical laws have translational and rotational spatial symmetry (relative to a cosmological coordinate system in which the description of the cosmic medium is invariant under spatial translations and rotations or relative to a local inertial coordinate system), the strong cosmological principle follows from the assumption that a model that has these symmetries completely describes the early universe at a single moment of time.

The strong cosmological principle is inconsistent with classical microphysics. In a uniform, unbounded, infinite distribution of classical particles, every particle is uniquely situated with respect to its neighbors, or even with respect to its nearest neighbor, because the number of pairs of nearest neighbors is countably infinite whereas the number of possible (real) values of the ratio between two nearest-neighbor separations is uncountably infinite. By contrast, a complete quantum description of a uniform, unbounded distribution of non-interacting particles specifies a number density and a set of single-particle occupation numbers for each kind of particle. The following discussion will bring to light other links between cosmology and quantum physics.

## A. Statistical cosmology

Non-uniform cosmological models that conform to the strong cosmological principle represent the distribution of mass, temperature, chemical composition, and other macroscopic variables by random functions of position. Each random function is characterized by an infinite set of probability distributions each of which is invariant under spatial translations and rotations. For example, the spatial distribution of mass at a particular moment is specified by the probability distribution of the mass density at a single point, the joint-probability distribution of the mass densities at two given points, the joint-probability distribution of the mass densities at three given points, and so on. The probability distribution of the mass density at the point $x = (x^1, x^2, x^3)$ is independent of $x$:

$$\Pr\{a \le \rho(x) \le a + da\} = p(a)da\,, \tag{1}$$

where $p(a)da$, the probability that the mass density lies in the interval $(a, a+da)$, is the same at all points. Similarly, the joint-probability distribution of the mass densities at two distinct points,

$$\Pr\{a \le \rho(x_1) \le a+da, b \le \rho(x_2) \le b+db\} = p(a,b;|x_1 - x_2|)dadb\,, \tag{2}$$

depends only on the distance $|x_1 - x_2|$ between the points but not on the direction of the vector joining them, and so on.

The probabilities that figure in a description that conforms to the strong cosmological principle can be *defined* in terms of spatial averages. For example, $\Pr\{a \le \rho(x) \le a+da\}$ is the mean, or spatial average, of the function

$$\delta(\rho(x); a, a+da) = \begin{cases} 1 \text{ if } \rho \in (a, a+da) \\ 0 \text{ otherwise} \end{cases} \,. \tag{3}$$

That is, the probability that the mass density at a given point lies in a given interval is the *fractional volume* (defined below) occupied by points at which the mass density satisfies this condition. Similarly, the joint probability that the mass densities at two points *x*, $x+y$ lie in given intervals is the fractional volume occupied by points *x* at which this condition, with *y* fixed, is satisfied; and so on.

More generally, let *I* denote a function that, like $\delta$ in (3), takes the value 1 at points where a given condition is satisfied and takes the value 0 at points where the condition is not satisfied. The probability that the condition is satisfied is the spatial average of *I*, or fractional volume of the (infinite) region occupied by points where the condition is satisfied. To define this spatial average, let *v*(*V*) denote the integral of *I* over a region of volume *V*. If the ratio *v*(*V*)/*V* approaches a limit as *V* increases indefinitely and if this limit doesn't depend on the shape or location of the regions that figure in the definition, we define it as the spatial average of *I*.

In more detail, assuming for simplicity that space is Euclidean, consider a hierarchy of progressively coarser Cartesian grids. Let $C_k^{(n)}$ denote the *k*th cell belonging to the grid of level *n*; *k* is an integer. The members of a cell belonging to level 0 of the hierarchy of grids are points in space. The members of cells belonging to level $n+1$ are cells of level *n*, and each cell of level $n+1$ contains 27 cells of level *n* and is centered on one of them. Let $\bar{I}(n,k)$ denote the average value of *I* in cell $C_k^{(n)}$. Let $\varepsilon_n$ denote the least upper bound of the absolute value of the differences $\left[\bar{I}(n,k)-\bar{I}(n,k')\right]$ for all pairs $k,k'$ and a fixed *n*. We now *stipulate* that $\varepsilon_n \to 0$ as $n \to \infty$. Then the $\bar{I}(n,k)$ for a given *n* approach a common limit as $n \to \infty$, which we identify with the spatial average of *I*. We take the stipulation that $\varepsilon_n \to 0$ as $n \to \infty$ to be part of the definition of statistical homogeneity.

If we identify point sets in space (at a given moment in cosmic time) with events in a sample space, and define the probability of an event as the fractional volume of the set of points where a specified condition is satisfied, the axioms of probability theory become true statements about point sets in physical space at a given moment of cosmic time, provided our description of the cosmic medium satisfies the strong cosmological principle (now taken to include the preceding stipulation about spatial averages). Thus the present definition of probability provides a *model* of axiomatic probability theory. The model goes beyond the axiomatic definition because it captures the intuitive notion of indeterminacy associated with probability. For example, the mass density at a given point is objectively indeterminate, because a complete description of the cosmic medium that complies with the strong cosmological principle doesn't contain the value of the mass density at that point.

It is instructive to compare the present definition of probability with the definitions that figure in standard accounts of kinetic theory and statistical mechanics. Maxwell and Boltzmann assumed that every molecule in a finite sample of an ideal gas is in a definite but unknown molecular state. They identified the probability of a molecular state with the fraction of molecules in that state. Gibbs [6] represented the macroscopic states of a system in thermodynamic equilibrium by probability distributions of (classical) microstates. He identified the probability of a range of microstates with the relative frequency of that range in an ensemble, a collection of imaginary replicas of the system, each in a definite microstate. But he did not – as some later authors have done – assume that the macroscopic system in a macrostate characterized by a probability distribution of microstates was actually in one of these microstates. As discussed below, the present approach, like Gibbs's, characterizes macrostates by probability distributions of microstates, and it does not assume that the system is in one of these microstates. On the contrary, the probability distribution that characterizes a macrostate characterizes it *completely*.

## *1. Statistical entropy*

In a probabilistic description of the cosmic medium, statistical entropy provides an objective (and, as Shannon showed) essentially unique measure of randomness. The statistical entropy *S* of a discrete probability distribution $\{p_i\}$ is given by [7]:

$$S(\{p_i\}) = -\sum_i p_i \log p_i \,. \tag{4}$$

Shannon proved that the three following properties of the function *S* define it uniquely. (i) *S* is non-negative and vanishes only when one of the $p_i$ is equal to 1. (ii) *S* takes its largest value, log *n*, just in case there are *n* non-vanishing probabilities $p_i$, each equal to 1/*n*. (iii) *S* is hierarchically decomposable, in the sense defined by (7) below. The last property plays a key role in the following discussion. To derive (7) from (4), think of the index $\alpha$ as the label of a macrostate consisting of microstates ($\alpha\, i$) whose probabilities $p_i^{(\alpha)}$ for a fixed value of $\alpha$ add up to $p^{(\alpha)}$:

$$p^{(\alpha)} = \sum_{i\in\alpha} p_i^{(\alpha)} \,. \tag{5}$$

The conditional probability $p_{i|\alpha}$ of the microstate *i*, given that it belongs to the macrostate $\alpha$, is defined by the identity

$$p_i^{(\alpha)} = p_{i|\alpha} p^{(\alpha)} \tag{6}$$

From (4) – (6) it follows that

$$S\left(\left\{p_i^{(\alpha)}\right\}\right) = S\left(\left\{p^{(\alpha)}\right\}\right) + \sum_{\alpha} p^{(\alpha)} S\left(\left\{p_{i|\alpha}\right\}\right). \tag{7}$$

That is, the statistical entropy of the probability distribution $\left\{p_i^{(\alpha)}\right\}$, often called the Gibbs entropy, is the sum of the coarse-grained statistical entropy $S\left(\left\{p^{(\alpha)}\right\}\right)$ and a *residual statistical entropy*, equal to the weighted average of the statistical entropies associated with the microstructures of the macrostates.

If the microstates refer to an isolated system, the Gibbs entropy is constant in time (in classical statistical mechanics on account of Liouville's theorem, in quantum statistical mechanics because microstates evolve deterministically). So if the coarse-grained statistical entropy increases with time, the residual statistical entropy must decrease at the same rate. Gibbs proposed a suitably defined coarse-grained entropy as the "analogue" of thermodynamic entropy; but, as discussed below, the universal validity of this interpretation has been challenged.

## *2. Continuous probability distributions*

To apply Shannon's definition (4) to a continuous probability distribution $\{p(x)\}$, one partitions the range of *x* into discrete intervals $\Delta_i$ and sets $p_i$ equal to the integral of $p(x)$ over $\Delta_i$. The value of $S(\{p_i\})$ depends on how one chooses the intervals $\Delta_i$. It increases without limit as the size of an interval shrinks to zero.

Boltzmann and Gibbs defined the statistical counterpart of entropy as an integral,

$$-\int \log p(x) \cdot p(x) dx,$$

where $p(x)$ is the probability per unit phase-space volume at a point $x$ in $\mu$- or $\Gamma$-space. In effect, they partitioned phase space into cells of equal volume. Since the value of $\log p(x)$ depends on how one chooses the unit of phase-space volume, the preceding formula does not assign a unique value to $S$. Like thermodynamic entropy, statistical entropy in Boltzmann's and Gibbs's theories is defined only up to an additive constant.

However, if we divide each cell in phase space into $k$ equal parts, the integral decreases by an amount $\log k$, so the difference between the values of $S$ for probability distributions defined on the same partition of phase space approaches a finite limit as the volume of a cell shrinks to zero. In quantum statistical mechanics, statistical entropy, as defined by (4), has a well-defined zero point, because bounded systems have discrete sets of possible quantum states.

### *3. Information*

Shannon used negative entropy (negentropy), the discrete counterpart of Boltzmann's $H$, as a measure of information. In the present context it is more convenient to define the information $I$ of a discrete probability distribution as the amount by which the distribution's statistical entropy $S$ falls short of the largest value of $S$ that is consistent with prescribed values for such quantities as the mean energy and the mean concentration of a chemical constituent [4]:

$$I \equiv S_{\max} - S \tag{8}$$

This definition is convenient because, as discussed below, in cosmological contexts the largest allowed value of the statistical entropy per unit mass may increase with time; so processes that generate statistical entropy may also generate statistical information.

$I$ is non-negative and vanishes when $S = S_{\max}$; a maximally random probability distribution has zero information. If we interpret statistical entropy as a measure of randomness or disorder, as is customary, we can interpret information as a measure of a deficiency of disorder, or order.

Like *S*, *I* is not uniquely defined for continuous probability distributions; it depends on how we partition the continuous space of events on which the distribution is defined into cells. But once the space has been partitioned into cells, *I*, unlike *S*, approaches a finite limit when we divide each cell into *k* equal cells and let *k* increase indefinitely. In classical statistical mechanics, acceptable phase-space grids have cells of equal phase-space volume. As the cell size approaches 0, *I* approaches a finite limit (whereas *S* blows up).

In quantum statistical mechanics, *S* and *I* are both well defined. Because a region of phase space whose volume *V* is much larger than the volume *v* of a quantum cell ($h$ in $\mu$-space, $h^N$ in $\Gamma$-space) contains $V/v$ quantum states, classical and quantum estimates of *I* agree whenever the classical estimate applies.

Finally, follows easily from its definition that information, like statistical entropy, is hierarchically decomposable:

$$I\left(\left\{p_i^{(\alpha)}\right\}\right) = I\left(\left\{p^{(\alpha)}\right\}\right) + \sum_{\alpha} p^{(\alpha)} I\left(\left\{p_{i|\alpha}\right\}\right) \tag{9}$$

where each occurrence of *I* is defined by the appropriate version of (8). The second term represents residual information. The first term represent information associated with what Wigner, in the passage quoted at the beginning of this paper, called externally imposed gross constraints. His suggestion that initial conditions are as random as the externally imposed gross constraints allow implies that the second term vanishes. One aim of this paper is to ground this suggestion in a historical account of the initial conditions that define macroscopic systems and their surroundings.

## B. The growth of order [4], [5]

Standard accounts of the early universe assume that at the earliest times when current theories of elementary particles apply, the cosmic medium was a uniform, uniformly expanding gas containing “a great variety of particles in thermal equilibrium ... .” [2, p. 528] At these early times the relative concentrations of particle kinds and the distributions of particle energies were characterized by maximally random probability distributions. And though the mass density and the temperature were changing rapidly, the rates of particle encounters and reactions were high enough to keep the relative concentrations of all particle species close to the equilibrium values appropriate to the instantaneous values of the temperature and the mass density.

As the universe expanded, both its rate of expansion and the rates of particle encounters decreased, but the latter decreased faster than the former. As a result, some kinds of equilibrium ceased to prevail, and the corresponding probability distributions ceased to be maximally random. For example, in the standard evolutionary scenario, which assumes that the cosmic microwave background is primordial, matter and radiation decoupled when their joint temperature fell low enough for hydrogen to recombine. Thereafter, the matter temperature and the radiation temperature declined at different rates. So while matter-radiation interactions tended to equalize the matter and radiation temperatures, generating entropy in the process, the cosmic expansion drove the two temperatures farther apart, generating information. Earlier, the relative concentrations of nuclides were frozen in when the rates of nuclear reactions became too slow relative to the rate of cosmic expansion to maintain their equilibrium values. Again, competition between the cosmic expansion and nuclear reactions generated both entropy and information.

Local, as well as global, gravitational processes tend to disrupt thermodynamic equilibrium. The first astronomical systems may have formed when the uniform distribution of mass and temperature became unstable against the growth of density fluctuations, or else when primordial density fluctuations reached critical amplitudes. But unlike isolated gas samples, newly formed self-gravitating systems did not settle into equilibrium states. For example, a gas cloud of stellar mass has negative heat capacity:

its temperature rises as it loses energy through radiation. As the cloud continues to radiate and contract, the disparities between its mean density and its mean temperature and those of and those of its surroundings steadily increase. Within the cloud, gradients of temperature and mass density become progressively more marked. In a mature star, nuclear reactions in the core gradually alter its chemical composition, thereby producing chemical inhomogeneity – another variety of disequilibrium.

### *1. Cosmology and the second law of thermodynamics*

I have argued that while local processes drive local conditions toward local thermodynamic equilibrium, the cosmic expansion and the contraction of self-gravitating astronomical systems drive local conditions away from local thermodynamic equilibrium, creating both information and entropy. This argument raises the questions: In what domain is entropy defined? And in what domain is the Second Law valid?

In classical thermodynamics, entropy is initially defined for systems in thermodynamic equilibrium. This definition is then extended to systems in local, but not global, thermodynamic equilibrium and to composite systems whose components are individually in global or local thermodynamic equilibrium. The Boltzmann-Gibbs-Shannon definition of statistical entropy allows us to extend the thermodynamic definition to any physical state defined by a probability distribution. Can the domain in which the Second Law is valid be likewise extended?

Gibbs showed that the statistical entropy of the probability distribution of classical microstates that defines a macrostate of a closed system is constant in time; the same is true of a probability distribution of classical microstates of a closed system. Boltzmann's $H$ theorem extends the Second Law to arbitrary non-equilibrium states of a sample of an ideal gas, but his proof of the theorem assumes that the residual statistical information (see Eq. 7) of the probability distribution that characterizes the sample's macrostate vanishes at all times. This assumption cannot be weakened. It isn't enough to assume that the residual information (i.e., information associated with molecular correlations) vanishes initially. Boltzmann's proof and Gibbs's proof of the constancy of a closed system's statistical entropy show that molecular interactions do not destroy single-particle information; they convert it into residual information; and Poincaré's theorem shows that in a closed system residual information residual information eventually makes its way back into single-particle information.

Boltzmann's assumption that residual statistical information is permanently absent exemplifies what van Kampen has called "repeated randomness assumptions." As discussed in more detail below, he argued that the *H* theorem exemplifies a wide class of theorems about stochastic processes that rely on versions of this assumption. I argue below that these assumptions, when valid, are justified by historical arguments. It follows that *H* theorems – extensions of the Second Law to macroscopic systems whose macrostates can be characterized by probability distributions of quantum states – are historical generalizations, depend on historically justified initial and boundary conditions.

Can the notion of entropy and the law of entropy growth be extended to self-gravitating systems? Consider a self-gravitating gas cloud. We can regard a sufficiently small region of the cloud as a gas sample in a uniform external gravitational field, and conclude from the Second Law that physical processes occurring within it generate entropy. If the cloud's large-scale structure is not changing too rapidly, we can then conclude that *local* thermodynamic equilibrium prevails. We can, if we wish, define the entropy of the cloud as the sum of the entropies of its parts. But this definition does not contain a gravitational contribution. There is no consensus that such a contribution exists and no widely accepted view of how it might be defined if it does exist. Presumably it would be non-additive, like the gravitational contribution to the cloud's energy.

The gravitational contribution to the energy of a self-gravitating cloud causes the cloud to behave very differently from a closed gas sample in the laboratory. Although heat flow, mediated by collisions between gas molecules, tends to reduce temperature differences between adjacent regions, the cloud does not relax into a state of uniform temperature. (Indeed, such a state doesn't exist for a cloud of finite mass.) Instead, the cloud evolves toward a state of *dynamical* (quasi-)equilibrium in which the gravitational force acting on each element balances the resultant of the pressure forces on the element's boundary. The virial theorem, applied to a self-gravitating system in dynamical equilibrium, implies that the system has negative heat capacity: its mean temperature increases as the cloud loses energy through radiation. As the cloud contracts, the internal gradients of temperature, density, and pressure become steeper. The non-gravitational contribution to the cloud's entropy decreases.

## *2. Initial conditions for macroscopic systems*

Every observer sees the world from his or her own point of view. An account of the universe and its evolution that incorporates the strong cosmological principle privileges none of these local perspectives. In a sense, it includes all of them. Yet it cannot contain or predict initial conditions that characterize individual physical systems, such as the Sun. To make such a prediction the cosmological account would have to contain a sentence like "Relative to a specified coordinate system, the object whose center of mass is at $\boldsymbol{x}$ at time $t$ has such-and-such properties." But it cannot contain such a sentence, because there is no preferred coordinate system. The cosmological account has no way of singling out individual systems.

On the other hand, every observer acquires information about individual systems – information that isn't in the perspectiveless cosmological account. Where does that information come from if it wasn't in the (supposedly) complete cosmological account to begin with? As Szilard [8] pointed out long ago, the entropy generated by an observational process more than offsets the information delivered by the observation. Part of the process consists in converting information in information-rich fuel sources (batteries, ATP molecules) into qualitatively new forms (improved estimates of the Sun's mass). So we can acquire an indefinite quantity of data that characterize the view from Earth. Our description realizes one of infinitely many possible descriptions, centered on different points in space. All these descriptions contain the same *statistical* information. The perspectiveless cosmological account contains *only* that information.

Every macroscopic system and its surroundings are characterized by probability distributions of microstate. Such a probability distribution is determined by its history, which began in the early universe. But the history of every macroscopic system also contains local contributions. In particular, the history of a system on a laboratory bench includes an account of how the system and its surroundings were prepared. I will argue below that macroscopic systems are rarely found in definite microstates (the argument is due to Zeh [32]) but may be, and often are, found or prepared in definite macrostates. This argument lies at the heart of the solution to the measurement problem described below.

## III. CHANCE IN THE MACROSCOPIC DOMAIN

In the 1880s Poincaré discovered that some orbits in self-gravitating systems are extremely sensitive to initial conditions. In a popular essay published twenty years later [9], he suggested that extreme sensitivity to initial conditions is the defining characteristic of what is now called deterministic chaos. The outcomes of such processes, Poincaré argued, are predictable in principle but unpredictable in practice.

Poincaré's first example was a cone initially balanced on its tip. Any external disturbance, no matter how small, causes the cone to topple in a direction that is unpredictable in practice. The following argument suggests that even in the absence of external disturbances and even if we neglect Heisenberg's uncertainty principle, the direction of fall is unpredictable in principle as well as in practice.

Classical mechanics represents the cone's possible microstates by points in a four-dimensional phase space whose coordinates are the elevation and azimuth of the cone's axis and the conjugate angular momenta. But according to the arguments of §II, the cone's initial macrostate is represented not by a single point in this phase space but by a probability distribution of possible microstates. This probability distribution must contain a finite quantity of information; its support must have finite phase-space volume, which in general will greatly exceed $\hbar^2$, the limit set by Heisenberg's uncertainty relation. If the support includes the point that represents the initial state of unstable equilibrium, it contains classical phase-space trajectories that correspond to all possible directions of fall, and the final orientation of the axis is unpredictable in principle. From a description of the experimental setup one could in principle infer the probability distribution of classical microstates that characterizes the cone's initial macrostate.

Classical mechanics would allow us to calculate the evolution of each microstate and hence to calculate the probability distribution of final microstates; and a description of the apparatus that records the final orientation of the cone's axis would allow us to partition this distribution into a distribution of macrostates. As discussed below, an analogous schema applies to quantum measurements.

The probability distributions that characterize the initial states of macroscopic systems depend on their history. Hence there can be no genuine laws about initial conditions, only historical generalizations. For example, macroscopic systems cannot usually be prepared in definite quantum states; but physicists have succeeded in preparing superconducting quantum interference devices (SQUIDs) in superpositions of macroscopically distinct quantum states. Again, macroscopic systems are usually microscopically disordered; but Hahn's spin-echo experiment showed that this is not necessarily the case.

Nevertheless, the argument of §II suggests that many natural processes have unpredictable outcomes. This conclusion, if correct, would bring the physicists' worldview closer not only to the worldview of ordinary experience, in which chance seems to play a major role, but also to that of biology. Chance plays a key role in evolution: genetic variation, which generates candidates for natural selection, has a random component, and the histories of individuals, populations, and species are strongly influenced by apparently unpredictable fluctuations in their environments. Some developmental processes – the immune response, visual perception, and some learning strategies, for example – likewise rely on processes with unpredictable outcomes [10].

### A. Equilibrium statistical mechanics

Most modern presentations of classical and quantum statistical mechanics follow the mathematical approach pioneered by J. W. Gibbs [6], in which probability distributions of an undisturbed *N*-particle system's microstates represent equilibrium macrostates and probability-weighted averages of microscopic quantities represent thermodynamic variables. Quantum statistical mechanics represents macroscopic states and variables in exactly the same way. Let the probability distribution $\{ p_k^{(\alpha)} \}$ of quantum states $k$ characterize a macrostate $\alpha$, and let $O$ be an observable. The statistical counterpart of $O$ is

$$\overline{O}^{(\alpha)} \equiv \sum_{k \in \alpha} p_k^{(\alpha)} \langle k | O | k \rangle = \mathrm{Tr}\,(\rho^{(\alpha)} O) \qquad (10)$$

where

$$\rho^{(\alpha)} = \sum_{k \in \alpha} | k \rangle p_k^{(\alpha)} \langle k | \qquad (11)$$

$\overline{O}^{(\alpha)}$ is the value of the macroscopic variable $\overline{O}$ in the macrostate $\alpha$. We discuss the physical interpretation of (10) in §III.C below.

The statistical counterpart of entropy is statistical entropy:

$$S^{(\alpha)} = -\sum_{k \in \alpha} p_k^{(\alpha)} \log p_k^{(\alpha)} \qquad (12)$$

Gibbs called his statistical descriptions "analogies." He showed that the canonical probability distribution,

$$p_k = e^{-E_k/\mathrm{k}T} \Big/ \sum_j e^{-E_j/\mathrm{k}T} \;, \qquad (13)$$

maximizes the statistical entropy, subject to the constraint that the mean energy has a prescribed value, represented by the thermodynamic variable $E$. The reciprocal of the Lagrange multiplier associated with this constraint is the absolute (Kelvin) temperature $T$. Gibbs showed that the thermodynamic identity $TdS = dE + PdV$ obtains in the statistical description based on the canonical distribution. H also showed that the statistical description based on the microcanonical distribution, in which the probability distribution is uniform over the $(6N - 1)$-dimensional region of phase whose points represent microstates with a given energy, mimics the thermodynamic description less successfully.

Some authors have tried to *derive* statistical mechanics from mechanics. These authors favor the microcanonical distribution, because they assume that an undisturbed macroscopic system is in a definite microstate, and hence has a definite energy. They replace Gibbs's averages over phase space (or, in quantum statistical mechanics, microstates) by time averages. Although this approach has generated interesting mathematics, it is limited to equilibrium statistical mechanics, and even in that context it relies on questionable ad hoc assumptions.

Other authors have sought to base statistical mechanics on an epistemic argument:

> [M]acroscopic observers, such as we are, are under no circumstances capable of observing, let alone measuring, the microscopic dynamic state of a system which involves the determination of an enormous number of parameters, of the order of $10^{23}$. ... [Thus] a whole *ensemble* of possible dynamical states corresponds to the *same* macroscopic state, compatible with our knowledge. [11]

As mentioned above, Gibbs proved that the statistical entropy of the canonical distribution exceeds that of any other distribution with the same mean energy. E. T. Jaynes [12], [13] made this theorem the cornerstone of an elegant formulation of statistical mechanics that interprets statistical entropy as a measure of missing knowledge. He argued that "statistical mechanics [is] a form of statistical inference rather than a physical theory." Its "computational rules are an immediate consequence of the maximum-entropy principle," which yields "the best estimates that could have been made on the basis of the information available." [11, p. 620] Gibbs, by contrast, interpreted statistical entropy not as a measure of missing knowledge – he did not identify the macroscopic system he was describing with a definite but unknown member of the ensemble that characterizes the system's thermodynamic state – but as an "analogue" of thermodynamic entropy. In the present account, the probability distributions that characterize a macroscopic system and its surroundings characterize them completely and objectively; they have nothing to do with what we know or don't know about the system and its surroundings; and statistical entropy is not a measure of human ignorance but of objective randomness.

Gibbs also proved that the canonical probability distribution characterizes subsystems of an extended isolated system characterized by a microcanonical probability distribution. That is, it characterizes a macroscopic system in a heat bath. Many modern authors (e.g., Schrödinger [14]) justify the canonical distribution on these grounds. The foundational problem then shifts its focus from the canonical (or grand canonical) distribution to the microcanonical distribution and becomes the problem of justifying the ergodic hypothesis.

The considerations of §II supply an objective version of Jaynes's maximum-entropy principle. Jaynes's principle encapsulates the sound methodological precept that a scientific description of a physical system should not outrun what scientists know, or can know, about the system. In the present account, the probability distribution that *objectively* characterizes a macrostate contains just the information created by the physical processes that shaped that macrostate. In particular, probability distributions that objectively characterize equilibrium states have zero residual information. Consequently, Jaynes's methodological precept normally leads to the same statistical description of equilibrium states as a historical physical argument based on an assumption about the primordial cosmic medium. But it can happen that an objective statistical description of a macrostate contains residual information that an observer is unaware of. In this situation, illustrated by Hahn's spin-echo experiment, a prediction based on the maximum-entropy principle fails.

### B. Non-equilibrium Statistical Mechanics

According to the present approach, a complete description of a macroscopic system's initial state contains just the information created by the system's history, including its method of preparation. This supplies a framework – but only a framework – for addressing what Van Kampen [15] has called "the main problem in the statistical mechanics of irreversible processes": What determines the choice of macrostates and macroscopic variables?

Van Kampen [15, 16] has also emphasized the role of "repeated randomness assumptions" in theories of stochastic processes:

> This *repeated randomness assumption* is drastic but indispensable whenever one tries to make a connection between the microscopic world and the macroscopic or mesoscopic levels. It appears under the aliases "Stosszahlansatz," "molecular chaos," or "random phase approximation," and it is responsible for the appearance of irreversibility. Many attempts have been made to eliminate this assumption, usually amounting to hiding it under the rug of mathematical formalism. [16, p. 58]

To derive his transport equation and his *H* theorem, Boltzmann had to assume (following Maxwell) that the initial velocities of colliding molecules in a closed gas sample are statistically uncorrelated. This assumption cannot hold for non-equilibrium states, because when the information that characterizes a non-equilibrium state decays, residual information associated with molecular correlations comes into being at the same rate. However, because a gas sample's reservoir of microscopic information is so large, molecular chaos may prevail to a good approximation for periods much shorter than the Poincaré recurrence time, which is typically much longer than the age of the universe.

Another approach to the problem of justifying repeated randomness assumptions starts from the remark that no gas sample is an island unto itself. Can the fact that actual gas samples interact with their surroundings justify the assumption that correlation information is permanently absent in a nominally closed gas sample? And if so, is it legitimate to appeal to environmental "intervention"?

Fifty years ago, J. M. Blatt [17] argued that the answer to both questions is yes. The walls that contain a gas sample are neither perfectly reflecting nor perfectly stationary. When a gas molecule collides with a wall, its direction and its velocity acquire tiny random contributions. These leave the single-particle probability distribution virtually unaltered, but they alter the histories of individual molecules, thereby disrupting multi-particle correlations. Blatt distinguished between states of true equilibrium, characterized (in the vocabulary of the present paper) by an information-free probability distribution, and quasi-equilibrium states, in which single-particle information is absent but correlation (or residual) information is present. With the help of a simple thought experiment, Blatt argued that collisions between molecules of a rarefied gas sample and the walls of its container cause an initial quasi-equilibrium state to relax into true equilibrium long before the gas has come into thermal equilibrium with the walls.

Earlier, Bergmann and Lebowitz [18] constructed and investigated detailed mathematical models of the relaxation from quasi-equilibrium to true equilibrium through external "intervention." More recently, Ridderbos and Redhead [19] have expanded Blatt's case for the interventionist approach. They constructed a simple model of the spin-echo experiment [20], in which a macroscopically disordered state evolves into a macroscopically ordered state. They argued that in this experiment (and more generally) interaction between a nominally isolated macroscopic system and its environment mediates the loss of correlation information.

Blatt noted that this "interventionist" approach "has not found general acceptance."

> There is a common feeling that it should not be necessary to introduce the wall of the system in so explicit a fashion. ... Furthermore, it is considered unacceptable philosophically, and somewhat "unsporting," to introduce an *explicit* source of randomness and stochastic behavior directly into the basic equations. Statistical mechanics is felt to be a part of mechanics, and as such one should be able to start from purely causal behavior. [17, p. 747]

The historical approach sketched in this paper directly addresses these concerns. It implies that statistical mechanics cannot be derived from quantum (or classical) microphysics, because, as discussed below, macroscopic systems cannot be idealized as closed systems in definite microstates. It assumes that they *can* be idealized as closed systems in definite *macro*states. And it anchors the statistical assumptions about the initial states of macroscopic systems and their environments on which statistical theories depend in a historical narrative based on a simple cosmological assumption.

Derivations of master equations that rely on decoherence ([21] and references cited there]) exemplify and significantly extend the interventionist approach. These derivations assume (unrealistically) that macroscopic systems are initially in definite quantum states but also (realistically) that they interact with random environments. Under these conditions, environmental interactions transfer information very effectively from the system to its surroundings. Microscopic information that is wicked away from the system disperses outward, eventually getting lost in interstellar and intergalactic space.

### C. The origin of irreversibility

Because the microstates of undisturbed systems evolve reversibly, a theory that assigns macroscopic systems (or the universe) definite quantum states cannot provide a framework for theories that distinguish in an absolute sense between the two directions of time. The historical approach sketched in this paper offers such a framework because by linking the initial states of macroscopic systems to states of the early universe it equips time with an arrow. Specifically, it predicts that macroscopic systems are *usually* embedded in random environments and that their initial states *usually* contain only information associated with the values of macroscopic variables. However, as the spin-echo experiment shows, macroscopic systems *can* be prepared in states that contain microscopic information.

### D. Quantum measurements

#### *1. Cosmological ensembles*

A description of the universe that comports with the strong cosmological principle should not pick out any particular macroscopic system. We therefore assume that macroscopic descriptions refer not to individual systems but to *cosmological ensembles* [4], [5] – infinite collection of identical systems distributed in a statistically uniform way in space. We interpret the density operator $\rho^{(\alpha)}$ in Eq. (10) as referring to such an ensemble:

Interpret the quantity Tr $(\rho^{(\alpha)}O)$ as the average value of the macroscopic variable $\overline{O}^{(\alpha)}$ in a cosmological ensemble. (14)

As in (10), $\overline{O}^{(\alpha)}$ is the value of $\overline{O}$, the macroscopic counterpart of the observable *O*, in the macrostate $\alpha$ characterized by the density operator $\rho^{(\alpha)}$.

Rule (14) resembles a rule given by Dirac [22, pp. 132, 133]:

Identify Tr ($\rho O$) with the average result of a large number of measurements of *O* when the system "is in one or other of a number of possible states according to some given probability law." (14*)

Dirac's rule (14*) generalizes his interpretation of the matrix element $\langle s|O|s\rangle$ as the average result of a large number of measurements of *O* when the system is in state *s*. Like the earlier rule, it treats measurement as a primitive concept. In contrast, rules (10) and (14) do not mention measurement; it is trivially true that $\overline{O}^{(\alpha)}$ is the result of measuring $\overline{O}$ when the system is in the macrostate characterized by $\rho$. Nor do (10) and (14) presuppose Dirac's interpretation of $\langle s|O|s\rangle$. As discussed in §D.2, (14) allows us to *derive* this interpretation from a modified form of von Neumann's account of an ideal measurement, and thus to bring quantum measurement into the domain of quantum statistical mechanics.

The density operator $\rho^{(\alpha)}$ refers to a cosmological ensemble whose members are all in the macrostate $\alpha$. Suppose $\alpha$ is the initial quantum state of Poincaré's cone, whose axis is as nearly vertical and as nearly stationary as Heisenberg's uncertainty relations allow. If the cone is undisturbed, its microstates evolve deterministically, and the density operator that represents its macrostate evolves into a density operator that represents a number of macroscopically distinguishable macrostates:

$$\begin{aligned}\rho &= \sum_{\alpha}\sum_{k} |\alpha k\rangle p_k^{(\alpha)} \langle \alpha k| \\ &= \sum_{\alpha} p^{(\alpha)} \sum_{k} |\alpha k\rangle p_{k|\alpha} \langle \alpha k|\end{aligned} \tag{15}$$

(As earlier, $\alpha k$ denotes the *k*th microstate of the macrostate of the macrostate $\alpha$, $p_k^{(\alpha)}$ denotes the probability of the microstate $\alpha k$, $p_{k|\alpha} = p_k^{(\alpha)} / p^{(\alpha)}$, and $p^{(\alpha)} = \sum_k p_{k|\alpha}$ .) Then

$$\text{Tr}\,(\rho O) = \sum_{\alpha} p^{(\alpha)} \sum_{i} p_{i|\alpha} \langle \alpha i|O|\alpha i\rangle = \sum_{\alpha} p^{(\alpha)} \overline{O}^{(\alpha)} \tag{16}$$

By (14), this is the ensemble average of the macroscopic counterpart of the observable *O*. In particular, if $P_\beta$ is the projector

$$P_\beta = \sum_k |\beta k\rangle\langle \beta k|\,, \tag{17}$$

$$\text{Tr}\,(\rho P_\beta) = p^{(\beta)} \tag{18}$$

Since $P_\beta$ takes the value 1 in microstates that belong to the macrostate $\beta$ and the value 0 in microstates that belong to other macrostates that figure in the definition (15) of $\rho$, $\text{Tr}\,(\rho P_\beta)$ is the fraction of ensemble members in the macrostate $\beta$. Eq. (18) says that this fraction is $p^{(\beta)}$.

Thus the ensemble interpretation accounts for the kind of macroscopic indeterminacy exemplified by Poincaré's thought experiment: To avoid giving special standing to a particular cone, and hence to its position in the universe, we think of it as a member of a cosmological ensemble. If the probability distribution of microstates that defines the cone's initial macrostate does not contain enough information to predict its final macrostate (defined by the orientation of the cone's axis), it evolves (deterministically) into a probability distribution that can be partitioned into sub-distributions that represent macroscopically distinguishable macrostates. A complete description of the experimental setup would allow one to assign a probability to each of these macrostates. This probability is its relative frequency in a cosmological ensemble. The outcome at a particular place is unpredictable because all members of a cosmological ensemble are on the same footing.

## *2. Quantum measurements*

Von Neumann's account of an ideal measurement [23] assumes that the combined system (measured object + measuring apparatus) is initially in a definite quantum state; that Schrödinger's equation governs the evolution of this state during a measurement; and that if the object is initially in an eigenstate of the measured observable, the measurement leaves it in that state. It follows from these assumptions that if the object is initially in a superposition of eigenstates of the measured observable, the final microstates of the combined system are superpositions of macroscopically distinguishable microstates, each associated with one of the measurement's possible outcomes (as predicted by the standard interpretation rule). These assumptions define an ideal measurement – or, more precisely, a "premeasurement," which is followed by a collapse of the superposition onto one of its components.

In a historical account of cosmic structure and evolution that comports with the strong cosmological principle, macroscopic systems cannot in general be idealized as being in definite quantum states. They can, however, be idealized as being in definite macrostates. The macrostate of a macroscopic system, such as a measuring apparatus, is defined by a probability distribution of microstates, which is determined by the system's history. Because the history of a macroscopic system stretches back to the early universe, the probability distribution that defines its macrostate contains only the information created by the information-generating processes, including experimental preparation, that the system has undergone. In general, the information that characterizes the initial state of a measuring apparatus is that associated with the apparatus's specifications. Now, quantum mechanics does not specify the phases of individual microstates. So if a macroscopic system's history has not created information about the relative phases of the microstates that belong to its macrostate, as will always be the case for a measuring apparatus, a complete description of the macrostate will contain no information about relative phases. They will be random.

Decoherence calculations, by contrast, begin by assuming, with von Neumann, that the combined system is in a definite quantum state. Interaction between the combined system (S+M) and an environment E *that is assumed* to be random transfers relative-phase information from the superposition that would represent the state of S+M in the absence of environmental interaction to the superposition that represents the state of S+M+E.

Suppose then that the measuring apparatus is initially in a definite macrostate, characterized by a probability distribution $\{ p_i \}$ of quantum states $|i\rangle_{\mathrm{M}}$. Assume that the initial state of the measured system S is a superposition, with coefficients $c_k$, of eigenstates $|k\rangle_{\mathrm{S}}$ of an observable *K*. As in von Neumann's account, a measurement of *K* produces an entangled quantum state of the combined system:

$$\left(\sum_k c_k |k\rangle_{\mathrm{S}}\right)|i\rangle_{\mathrm{M}} \to \sum_k c_k |k\rangle_{\mathrm{S}} |j(k,i)\rangle_{\mathrm{M}} \equiv \sum_k c_k |k, j(k,i)\rangle \tag{19}$$

Here $j(k,i)$ labels a quantum state of the measuring apparatus. The argument *k* indicates that it is correlated to the eigenvalue *k* of the system S.

Let *O* denote an observable that refers to the combined system S + M. By (14), the ensemble average of the macroscopic counterpart of the observable *O* in the final state is:

$$\begin{aligned}\mathrm{Tr}\,(\rho O) &= \sum_i p_i \left[ \left\{ \sum_{k'} c_{k'}^* \langle k', j(k',i)| \right\} O \left\{ \sum_k c_k |k, j(k,i)\rangle \right\} \right] \\ &= \sum_i p_i \sum_{k',k} c_{k'}^* c_k \langle k', j(k',i)|O|k, j(k,i)\rangle \end{aligned} \tag{20}$$

Now, as Bohr emphasized (and experimental practice confirms), states of the measuring apparatus are classical states, completely specified by the values of classical variables; a complete description of the probability distribution $\{p_i\}$ that characterizes the measuring apparatus's initial state therefore contains no information about the relative phases of the quantum states *i*. As Bohm [22] pointed out long ago, they have random phases. What about the relative phases of quantum states $|j(k,i)\rangle_{\mathrm{M}}$, $|j(k',i)\rangle_{\mathrm{M}}$ belonging to different pointer states $k$, $k'$ of the measured observable? Decoherence theory shows that interaction between a macroscopic system and a random environment randomizes the relative phases of such states on very short time scales. Consequently the relative phases of off-diagonal matrix elements $\langle k', j(k',i)|O|k, j(k,i)\rangle$ are randomly distributed between 0 and $2\pi$. Since the number of microstates $i >> 1$, the result of averaging $\langle k', j(k',i)|O|k, j(k,i)\rangle$ over the probability distribution $\{p_i\}$ is close to zero:

$$\sum_i p_i \langle k', j(k',i)|O|k, j(k,i)\rangle = 0 \quad (k \neq k') \tag{21}$$

So the right side of (20) reduces, in an excellent approximation, to

$$\sum_k |c_k|^2 \sum_i p_i \langle k, j(k,i)|O|k, j(k,i)\rangle = \sum_k |c_k|^2 \overline{O}^{(k,k)}, \tag{22}$$

where $(k, k)$ labels the macrostate of the combined system in which the object is in the quantum state *k* and the apparatus is in the correlated pointer state.

As in the discussion following Eq. (16), we conclude that in each member of the cosmological ensemble of combined systems that represents the outcome of the measurement, the object is in an eigenstate of the measured observable and the apparatus is in the correlated pointer state. Moreover, the fraction of ensemble members in the state $(k, k)$ is $|c_k|^2$, as Born's rule prescribes.

*2a) Quantum measurements generate coarse-grained entropy.*

In the preceding description of an ideal measurement, the same probability distribution $\{p_i\}$ of microstates characterizes the initial and final states of the combined system, so its statistical entropy (the Gibbs entropy) doesn't change. But the coarse-grained entropy associated with the probability distribution of macrostates of the combined system increases from 0 (= log 1) to $-\sum_j |c_k|^2 \log |c_k|^2$. This gain is compensated by an equal loss of residual entropy. The loss of residual statistical entropy results mainly from the fact that the preceding idealized account allots fewer microstates to each of the final macrostates than to the initial macrostate. For example, if the initial macrostate contains $n$ microstates, with $n ? 1$, its statistical entropy is approximately log $n$. In the final state, the $n$ microstates are distributed among the final macrostates; so their residual entropy is approximately $\sum_\alpha p_a \log n_\alpha$, where $n_\alpha = p_\alpha n$. The total (fine-grained) statistical entropy of the final probability distribution is thus $-\sum_\alpha p_a \log p_a + \sum_\alpha p^{(\alpha)} \log n_\alpha = \log n$.

*2b) Decoherence and measurement*

Many previous attempts to apply quantum mechanics to the measurement process have implicitly or explicitly invoked the cancellation of matrix elements that connect macroscopically distinguishable quantum states in sums like (20). For example, Bohm [22] argued that in an idealized Stern-Gerlach experiment, the conditions for a good measurement require the phase of the silver atom's center-of-mass wave function to vary by an amount much greater than 2π over each region of the detector where the amplitude of the wave function is appreciable. Decoherence theory shows how interaction between a measuring apparatus and a random (internal or external) environment in effect randomizes the relative phases of quantum states belonging to different pointer states. It also supplies estimates of decoherence times for particular models. The preceding account of measurement appeals to the same process at a crucial step in the argument, the step leading to (21).

The preceding account supplements (and differs from) accounts based on decoherence in four significant ways:

(i) Decoherence theory applies Schrödinger's equation to an undisturbed system consisting of the measured system, the measuring apparatus, and a portion of the environment. In the present account the measuring apparatus is initially in a definite classical state, characterized by a historically defined probability distribution of quantum states.

(ii) Decoherence theory *postulates* that the (external or internal) environment has a random character. This is a crucial assumption. Its *operational* meaning is clear, but its physical meaning in the context of measurement theory and, more generally, in quantum statistical mechanics, is unclear. In the present account randomness has a definite and objective meaning: in a particular context it is measured by the statistical entropy of a probability distribution "descended" from the probability distributions that (objectively) characterize the primordial cosmic medium.

(iii) Because the present account explains environmental randomness, it predicts that quantum measurements are irreversible.

(iv) Decoherence theory explains why local measurements cannot exhibit the effects of interference between macroscopically distinguishable quantum states. But, as Joos and Zeh emphasized in a seminal paper,

> The use of the local density matrix allows at most only a partial derivation of classical concepts for two reasons: it already assumes a local description, and it presupposes the probabilistic interpretation leading to the collapse of the state vector at some stage of a measurement. ... The difficulties in giving a *complete* derivation of classical concepts may as well signal the need for entirely novel concepts [25].

In particular, decoherence theory does not solve what Schlosshauer [21] calls "the problem of outcomes;" it does not predict that measurements have definite outcomes. The present account does make this prediction. In a sense it also explains why quantum measurements have indeterminate outcomes: a complete description of the universe that comports with the strong cosmological principle cannot tell us *where* a measurement outcome occurs.

### E. Quantum measurements and general relativity

A quantum measurement can result in an unpredictable change in the local structure of spacetime. For example, it can cause an unpredictable displacement of a massive object.[1] This has been said to demonstrate the need for "treating the spacetime metric in a probabilistic fashion – i.e., [for] quantizing the gravitational field ..." [26]

Now the preceding account of a quantum measurement does not predict where each of the possible outcomes of a quantum measurement occurs. Instead it predicts what fraction of the membership of a cosmological ensemble experiences each possible outcome.

The present description of the cosmic medium (§II) is likewise probabilistic. Although the stress-energy tensor is a classical field, it is also a realization – indeed *the* realization – of a probabilistic description that does not privilege any position or direction in space. The indeterminacy of quantum measurement outcomes is consistent with this probabilistic description of the structure and contents of spacetime. We cannot predict the outcome of a quantum measurement because we cannot know where, in an absolute sense, it occurs. The structure of spacetime after a measurement has taken place reflects

[1] Such a displacement affects the structure of spacetime only locally, because it does not change the center of mass of the self-gravitating system in which the measurement occurs. Its effects therefore diminish rapidly with distance and are eventually lost in the noise associated with random fluctuations in the mass density of the universe.

the fact that a measurement has occurred but supplies no information about where it has occurred.

There are, of course, strong reasons for seeking a quantum theory of gravity. The preceding argument merely shows that if the historical account of initial conditions sketched in the preceding pages is correct, quantum mechanics and general relativity do not actually clash in the domain where they both hold, at least to an excellent approximation. The reason is that contact between them is mediated by statistical macrophysics, whose assumptions, as I have argued, are anchored in an account of cosmic evolution.

## IV. THE INTERPRETATION OF QUANTUM MECHANICS

Classical physics extends and refines intuitive and commonsense notions about the external world. As Dirac [22, preface to the first edition, 1930] pointed out, the formalism of quantum mechanics describes a mathematical world we can neither picture nor adequately translate into ordinary language. How is that world related to the world that classical physics describes? Although this question falls outside the domain of physics as such, an acceptable answer must comport with the rule that links the mathematical formalism of quantum mechanics to the results of possible measurements. And different versions of this rule have suggested different answers.

Dirac's version of the rule identifies the matrix element $\langle s|O|s\rangle$ with the average result of a large number of measurements of an observable $O$ when the measured system is in the state $s$. It is unambiguous, and it has proved adequate for all practical purposes. However, Dirac's exposition of quantum mechanics stops short of describing the measurement process itself.

Von Neumann's theory of measurement attempted to fill this gap. Von Neumann laid down necessary conditions for an ideal measurement. Assuming that the combined system is undisturbed and is initially in a definite quantum state, he concluded that the combined system evolves into a superposition of macroscopically distinguishable states, each of which represents one of the measurement's possible outcomes. To reconcile this result with Dirac's rule (and with experiment), he postulated that the predicted superposition promptly collapses onto one of its components.

Some attempts to avoid the need for such a postulate without altering the theory's formal structure appeal to methodological or philosophical considerations. Some prominent examples:

– Heisenberg [25] argued that state vectors do not describe objective physical states. Instead they describe a "potential reality" which measurements "actualize." Some contemporary physicists – for example, Gottfried and Yan [28, pp. 40, 574] – have embraced this view.

– Wigner [29] argued that "the function of quantum mechanics is not to describe 'reality,' whatever this term means, but only to furnish statistical correlations between subsequent observations. This assessment reduces the state vector to a calculational tool ..." Peres [30] agreed: "In fact, there is no evidence whatsoever that every physical system has at every instant a well defined state [whose] time dependence represents the actual evolution of a physical process."

– Everett [31] hypothesized that the universe is in a definite quantum state whose evolution is governed by Schrödinger's equation (or a linear generalization thereof). Every quantum measurement then creates a non-collapsing superposition of equally real quantum states. Proponents of this approach hope that the world of classical physics and experience will fit into this proliferating tree-like structure of universe-states, but they have not yet succeeded in showing precisely how.

– Zeh [30] argued that the quantum states of macroscopic systems, including the combined system in a quantum measurement, are invariably entangled with states of their environment. He argued that macroscopic systems, including large molecules, acquire classical properties through their interaction with a random environment, a process later dubbed "decoherence." Diverse theories that rely on this process [33] assume that quantum mechanics applies to macroscopic systems, including measuring apparatuses and their immediate environment, but not *necessarily* to the universe as a whole. Such theories explain why interference effects between macroscopically distinguishable states of a macroscopic system initially in a definite quantum state quickly become unobservable. They also explain the observed absence of transitions between macroscopically distinguishable states of large molecules (superselection rules) [30]. But they do no explain why quantum measurements have definite outcomes.

– Zurek [34], one of the founders of decoherence theory, has argued that quantum mechanics needs a cosmological postulate, namely, that "the universe consists of quantum systems, and ... that a composite system can be described by a tensor product of the Hilbert spaces of the constituent systems." Decoherence-based theories also need to assume that macroscopic systems are embedded in random environments. An account of cosmic evolution along the lines sketched in this paper would supply these needs.

In this paper I have argued that classical physics is neither a province nor a presupposition of quantum mechanics. Systems governed by classical laws are distinguished from systems governed by quantum laws by their histories, which determine the initial conditions that define them. These initial conditions take the form of probability distributions of quantum states. Probability distributions that assign comparable probabilities to a very large number of quantum states (i.e., probability distributions with broad support in Hilbert space) characterize systems governed by classical laws. Quantum laws govern systems to which one can assign a definite quantum state. Thus quantum physics and classical physics correspond to limiting cases of the historically determined initial conditions that define physical systems.

What is distinctive about this characterization of macroscopic states is not its mathematical form, which is old and familiar, but the interpretation of the probability distributions of quantum states that figure in it. I have argued that these probability distributions are descended from probability distributions that objectively describe states of the early universe; and that, in virtue of the strong cosmological principle, they belong to an objective description of the universe and its evolution. This interpretation offers an objective answer to van Kampen's "fundamental question: How is it possible that such macroscopic behavior exists, governed by its own equations of motion, regardless of the details of the microscopic motion?" [16, p. 56] It also offers an objective interpretation of quantum indeterminacy that contrasts with Heisenberg's view that the unpredictability of quantum-measurement outcomes results from "*uncertainties* concerning the microscopic structure of the [measuring] apparatus ... and, since the apparatus is connected with the rest of the world, of the microscopic structure of the whole world.” [25, pp. 53-4, my italics]

Mainstream interpretations of current physical theories offer a picture of the physical world in which the outcomes of physical processes other than quantum measurements and measurement-like processes are predictable in principle. The same physical theories, interpreted in light of the strong cosmological principle, suggest a qualitatively different picture of the physical world: one in which indeterminacy obtains at all levels of description. In this picture the universe itself is the unique realization of a statistical description that initially contains little or no statistical information. Gravitational processes – the cosmic expansion, the growth of density fluctuations in the cosmic medium, the contraction of self-gravitating systems – subsequently create statistical information, but the statistical information per unit mass remains far smaller than its largest possible value. Thus, as interventionist statistical theories of irreversibility and decoherence theories of the quantum-to-classical transition assume, the structure of the universe is orderly on macroscopic scales but random on microscopic scales. And because the initial states of macroscopic systems usually (though not necessarily and not always) contain little or no microscopic information, we can expect the outcomes of processes that depend sensitively on initial conditions to be objectively unpredictable in many cases.

The historical account of initial conditions sketched in this paper connects the physical laws that prevail at different levels of description but establishes no clear order of precedence among them. The strong cosmological principle is the source of indeterminacy at the macroscopic and astronomical levels, but it presupposes quantum mechanics. The strong cosmological principle also enables general relativity’s classical description of spacetime in the large to coexist peaceably with quantum indeterminacy (§III.E). Macroscopic laws of motion, which involve a small number of macroscopic variables, depend on cosmology, because macroscopic systems are defined by probability distributions that contain little or no microscopic information and are rooted in the probability distributions that characterize the early universe. At the same time, the predictions of macroscopic laws depend on the fact that quantum laws govern the microstructure of macroscopic systems.

The interconnectedness of the three levels of description allows macroscopic measuring devices to probe all three levels. But measurement does not play a privileged role in the present description. Nor does this description refer, implicitly or explicitly, to human knowledge. Thus although it assigns a central role to chance, it is consistent with Planck's and Einstein's view that physical theories describe observer-independent mathematical regularities behind the phenomena.

## ACKNOWLEDGMENTS

Anthony Aguirre and Silvan S. Schweber separately offered detailed, insightful, provocative, and very useful criticisms of several drafts of this paper. This version owes a great deal to their efforts (though it does not reflect their views on substantive issues addressed therein). I am deeply indebted to both of them.